\documentclass[journal]{IEEEtran}
\usepackage{amsmath,amsfonts}
\usepackage{algorithmic}
\usepackage{algorithm}
\usepackage{array}
\usepackage[caption=false,font=normalsize,labelfont=sf,textfont=sf]{subfig}
\usepackage{textcomp}
\usepackage{stfloats}
\usepackage{url}
\usepackage{verbatim}
\usepackage{graphicx}
\usepackage{cite}
\usepackage{booktabs} 
\usepackage{hyperref} 
\usepackage{listings}

\begin{document}

\title{Toward Generalized Autonomous Agents: A Neuro-Symbolic AI Framework for Integrating Social and Technical Support in Education}

\author{Ryan Hare,~\IEEEmembership{Student Member,~IEEE,}
        and Ying Tang,~\IEEEmembership{Senior Member,~IEEE}
\thanks{R. Hare and Y. Tang are with the Department of Electrical and Computer Engineering, Rowan University, Glassboro, NJ, 08028 USA (e-mail: harer6@rowan.edu; tang@rowan.edu).}
\thanks{This work was supported in part by the Rowan University-Rutgers-Camden Board of Governors, Google Research, and NSF Grants \#2121277 and \#1913809.}}

\markboth{IEEE SYSTEMS, MAN, AND CYBERNETICS MAGAZINE,~Vol.~XX, No.~X, MONTH~2024}%
{Hare \MakeLowercase{\textit{et al.}}: A Neuro-Symbolic Multi-Agent Framework}

\IEEEpubid{0000--0000/00\$00.00~\copyright~2024 IEEE}

\maketitle

\IEEEpubidadjcol

\IEEEPARstart{O}{ne} of the enduring challenges in education is how to empower students to take ownership of their learning by setting meaningful goals, tracking their progress, and adapting their strategies when faced with setbacks. Research has shown that this form of leaner-centered learning is best cultivated through structured, supportive environments that promote guided practice, scaffolded inquiry, and collaborative dialogue. 

In response, educational efforts have increasingly embraced artificial-intelligence (AI)-powered digital learning environments, ranging from educational apps and virtual labs to serious games. These apps and platforms are intentionally designed to teach academic content in engaging ways \cite{bachare}, and to provide more personalized guidance and support \cite{franzwa}. However, most of these efforts focus on developing intelligent tutoring systems (ITSs) that mimic direct teacher-student interaction by monitoring student learning behavior and providing corrective feedback. While effective for structured learning tasks, ITSs often treat learning as an isolated activity and overlook the social and collaborative dimensions essential for deeper understanding. Furthermore, ITSs are often tied to domain-specific learning content, which makes them costly and time-consuming to adapt to new subjects, instructional contexts, or diverse learner profiles—even though the underlying tutoring logic may, in principle, be generalizable.

Recent advances in large language models (LLMs) and neuro-symbolic systems, on the other hand, offer a transformative opportunity to reimagine how support is delivered in digital learning environments. LLMs are enabling socially interactive learning experiences and scalable, cross-domain learning support that can adapt instructional strategies across varied subjects and contexts \cite{wang2024parallel}. With the ability to engage in open-ended, human-like dialogue, LLMs introduce the potential for socially grounded learning experiences, enabling students to interact with AI agents in ways that mirror peer collaboration and guided discussion. In parallel, neuro-symbolic AI—which integrates the adaptability of neural networks with the structure and interpretability of symbolic systems—provide new avenues for designing these agents that are not only adaptive but also scalable across domains. 

Based on these remarks, this paper presents a multi-agent, neuro-symbolic framework designed to resolve the aforementioned challenges. The framework assigns distinct pedagogical roles to specialized agents: an RL-based 'tutor' agent provides authoritative, non-verbal scaffolding, while a proactive, LLM-powered 'peer' agent facilitates the social dimensions of learning. While prior work has explored such agents in isolation, our framework’s novelty lies in unifying them through a central educational ontology. As a backbone, the ontology ensures semantic consistency in instructional intent and agent behavior across subject areas, enabling the transfer of learning support strategies without being constrained by domain-specific content. Our system is designed to advance the next generation of AI-driven educational tools that are scalable to new topics, engaging, and learner-centered. In this article, we detail the architecture and implementation of the novel multi-agent educational framework. Through case studies in both college-level and middle school settings, we demonstrate the framework’s adaptability across domains. We conclude by outlining key insights and future directions for advancing AI-driven learning environments.

\section{Introduction}
The central goal of modern education is to cultivate learner agency—the capacity for students to set meaningful goals, monitor their progress, and adapt their strategies in a self-directed manner. Digital learning environments, from serious games to virtual labs, have emerged as powerful platforms for fostering this agency, offering engaging, hands-on practice that can be tailored to individual learning pathways. Yet, while these tools hold immense promise, the artificial intelligence designed to support learners within them often falls short, creating a significant gap between technological potential and pedagogical reality. The challenge lies in developing AI systems that provide not just instructional content, but a holistic support structure that is scalable, reliable, and acknowledges the fundamentally social nature of learning.

\begin{figure}
	\includegraphics[width=\columnwidth]{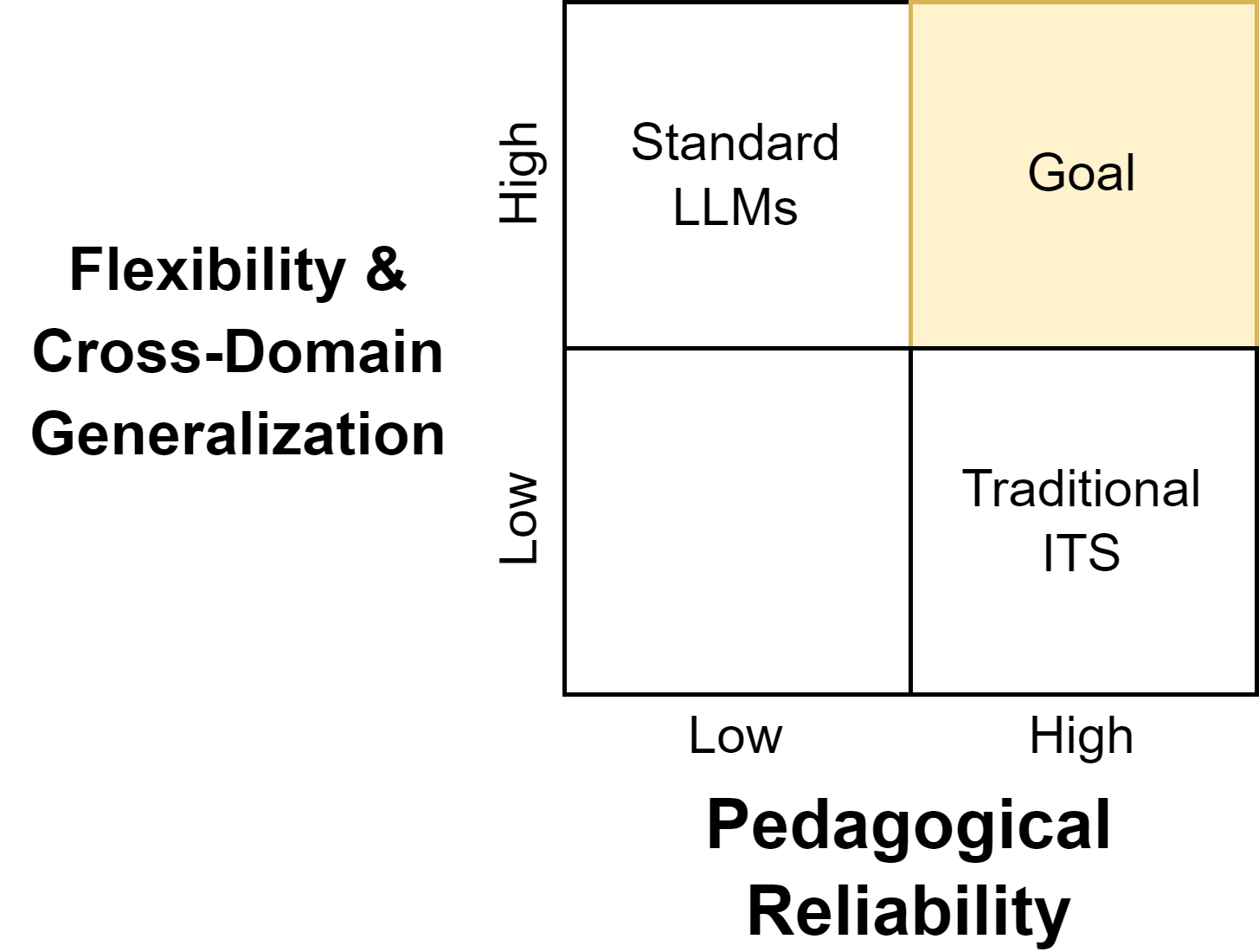}
	\centering
	\caption{Standard LLMs are decent tools, but despite recent developments, they have many issues in the education space. ITSs, meanwhile, are time- and resource-intensive to scale or apply to new topics. The goal is to achieve both scalability and educational effectiveness.}
	\label{fig:goals}
\end{figure}

This challenge has led to a critical issue in the field \cite{cui2023metaedu}, defined by a trade-off between two dominant but flawed approaches, as shown in Fig. \ref{fig:goals}. On one side is the generalizability crisis with traditional Intelligent Tutoring Systems (ITSs). ITSs are often built on hand-crafted expert rules, and these systems offer high pedagogical integrity and predictable behavior. This strength, however, comes at a cost: their architecture is notoriously difficult to scale and generalize. An ITS designed for university-level engineering cannot be repurposed for middle-school biology without a complete, resource-intensive re-engineering effort, rendering true scalability across the educational landscape prohibitively expensive \cite{murray2003, dermeval2018}.

On the other side, modern conversational AI has shown promise for educational support. The rise of Large Language Models (LLMs) presents a compelling alternative, offering remarkable conversational fluency and flexibility. However, LLMs are heavily dependent on prompting, available information, and context. When deployed without constraints, LLMs are prone to factual "hallucination," lack a persistent model of a learner's knowledge, and have poor inherent understanding of pedagogical strategy \cite{zhang2023hallucination}. Entrusting a student's learning to an unconstrained LLM can be dangerous, leading to conversations that could vary from inappropriate to incorrect, and ultimately having a negative impact on a student's education \cite{ref2, ref22}.

\begin{figure}
	\includegraphics[width=\columnwidth]{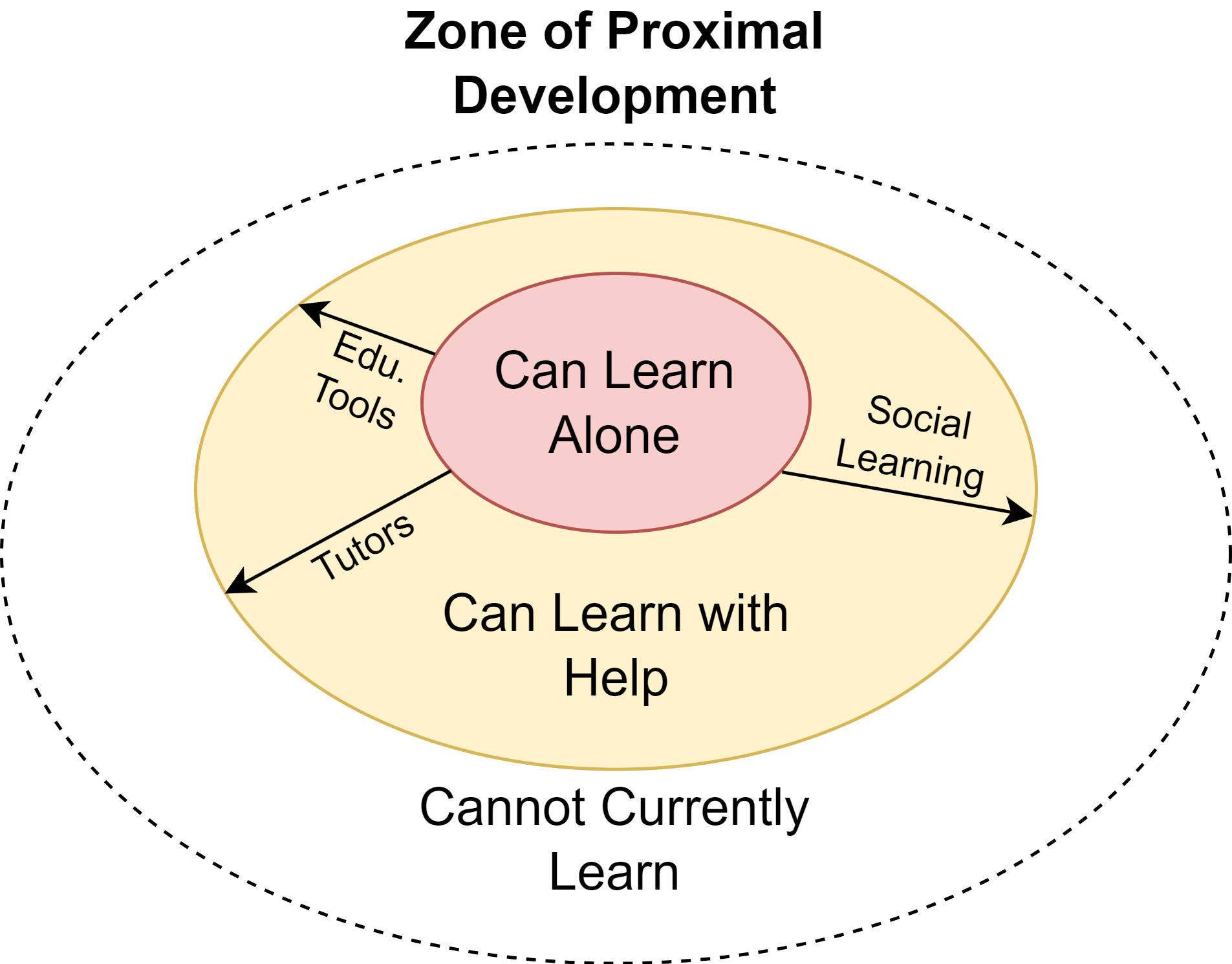}
	\centering
	\caption{Zone of Proximal Development. In this case, educational tools, tutors, and social learning all contribute to expanding what a student is able to learn, beyond what they can do independently.}
	\label{fig:ZPD}
\end{figure}

In addition to the issue of generalization, most current systems also have a pedagogical issue: they treat learning as a solitary task. When looking at decades of established learning science, it is clear that learning is as much a social task as it is an individual one. Sociocultural theories, particularly Vygotsky’s work on the Zone of Proximal Development (ZPD) \cite{vygotsky1978}, agree that learning is a fundamentally social process. Optimal learning occurs when a student receives not only "vertical" support from a More Knowledgeable Other (MKO), like a teacher providing expert scaffolding, but also "horizontal" support from peers through collaboration, discussion, and mutual inquiry (Fig. \ref{fig:ZPD}). Concepts like the protégé effect, where a student learns by preparing to teach a concept to someone else, further underscore the power of social interaction \cite{chase2014}. Despite this, the vast majority of AI in education research has focused exclusively on emulating the MKO, creating a significant social learning gap in the design of digital tools.

To create truly effective and holistic pedagogical agents with AI, we must design systems that address all three challenges (generalizability, effectiveness, and the social gap) simultaneously. The path forward lies not in choosing a side in the current impasse, but in the synergistic integration of paradigms. This points directly toward neuro-symbolic architectures, a burgeoning area of AI research focused on combining the pattern-recognition strengths of neural networks with the structured reasoning of symbolic systems like knowledge graphs (KGs) \cite{pan2024unifying}. This approach is gaining significant traction for enhancing LLM reasoning by grounding it in factual knowledge, with methods ranging from using Graph Neural Networks for retrieval-augmented generation (RAG) \cite{mavromatis2024gnn} to employing KGs to directly constrain the LLM decoding process \cite{pan2024unifying, luo2023reasoning}.

Drawing on this principle, this paper introduces a concrete architectural blueprint for a multi-agent, neuro-symbolic framework designed to overcome the limitations of current systems. Our framework addresses the three crises head-on by decomposing the problem: a central Educational Ontology acts as the symbolic backbone, providing a standardized structure to solve the scalability crisis while grounding the agents in verified knowledge to support educational effectiveness; a reinforcement learning-based Tutor Agent provides the adaptive, non-verbal scaffolding characteristic of an MKO; and a proactive, LLM-powered Peer Agent is explicitly designed to fill the social learning gap. This article details this framework, providing a conceptual and practical guide to its architecture and application as a step toward the next generation of autonomous AI tutors.

\section{Multi-Agent Tutoring Framework}\label{sec:method}
To address the challenges identified in the introduction, namely generalizability, educational effectiveness, and the social learning gap, our proposed solution is a multi-agent framework with each component serving to address the challenges. The core of our framework is the idea that a single AI cannot adequately solve these disparate problems. Instead, a comprehensive solution requires decomposing the complex task of tutoring into distinct pedagogical functions handled by specialized agents. Additionally, to support generalizability, our approach is neuro-symbolic at its core, combining the logical, verifiable structure of a symbolic knowledge base with the adaptive, generative power of neural agents.

\begin{figure*}
	\includegraphics[width=\textwidth]{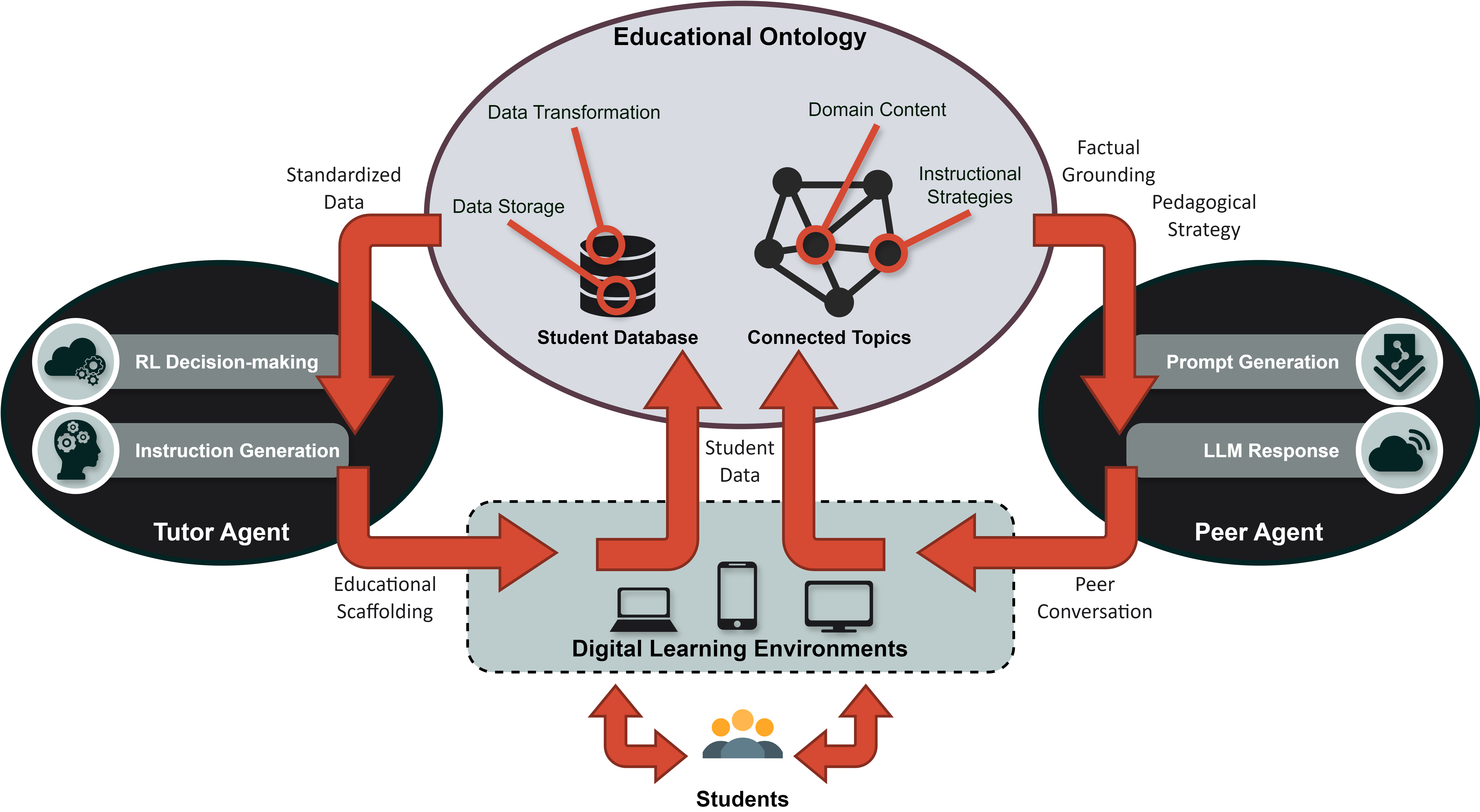}
	\centering
	\caption{Proposed system architecture. Raw interaction data are collected from the digital learning environment and transformed using the educational ontology, giving standardized data for use by both the peer and tutor agents.}
	\label{fig:PING}
\end{figure*}

The main goal of our framework is to unify specialized agents under a single, coherent architecture, as shown in Fig. \ref{fig:PING}. This architecture is designed for cross-domain applicability. And while other research has explored reinforcement learning for scaffolding or LLMs for educational dialogue, these efforts often remain isolated within their specific domains. Our contribution is the design of a system where the core logic can be shared across different subjects, learning contexts, and student populations \cite{ref12}. This is achieved by assigning distinct roles:
\begin{itemize}
	\item \textbf{Tutor Agent}: Provides the authoritative, non-verbal scaffolding of an expert.
	\item \textbf{Peer Agent}: Facilitates the collaborative, social dimensions of learning.
	\item \textbf{Educational Ontology}: Provides cross-domain commonality, acting as a data transformer and knowledge base that enables the same agents to operate effectively in any environment it describes.
\end{itemize}

\subsection{Educational Ontology}
The heart of our framework, and the direct answer to the generalizability and effectiveness crises, is a specialized knowledge graph we term the Educational Concept Ontology. This ontology is far more than a passive database; it is an active, multi-purpose component that serves as the system's central nervous system. It fulfills three roles:

\subsubsection{Enabling Generalizability}
The ontology is what makes our framework generalizable. It achieves this by creating an abstraction layer between the learning environment and the AI agents. Domain-specific knowledge (e.g., concepts like "Long Division" or "Trigonometry" for mathematics, their prerequisites, and associated learning materials) is encoded exclusively within the ontology files. The AI agents, in contrast, are programmed with domain-agnostic logic; they understand how to interpret the structure of the ontology (e.g., how to follow an is\_prerequisite\_for relationship) but have no hard-coded knowledge of its content. This separation is key. To deploy the entire system in a new subject, one simply provides a new ontology file. This addresses the high re-engineering cost that define the scalability crisis of traditional ITS.

\subsubsection{Abstracting and Transforming Data}
The ontology translates raw, heterogeneous data from any learning environment, $D_{raw}$, into a standardized, pedagogically meaningful student state vector, $S_t$. This mapping, represented conceptually as $S_t = f_{map}(D_{raw}|Ontology)$, is a function that transforms raw incoming data (e.g., mouse clicks, time on task, specific errors) and into high-level metrics (e.g., concept proficiency, frustration, engagement). Because both AI agents operate on this standardized state vector, their core logic remains entirely independent of the specific game or environment they are connected to.

\subsubsection{Informing Agents and Grounding LLMs}
The ontology serves as the system's "pedagogical conscience," providing a verifiable source of truth curated by human experts. For the Peer Agent, this repository of facts, relationships, and pedagogical strategies is used to guide and constrain the LLM's reasoning process. This directly supports educational effectiveness by ensuring the agent's dialogue is grounded in verified knowledge, preventing the factual "hallucinations" and pedagogical missteps characteristic of unconstrained LLMs.

Section \ref{sec:implementation} provides practical details on how the educational ontology is implemented to achieve these three functions.

\subsection{Tutor Agent}
The Adaptive Tutor Agent is our framework's embodiment of the "more knowledgeable other", designed to provide the structured, non-verbal scaffolding essential for effective learning. In contrast to the conversational Peer Agent, the Tutor operates as a silent, background optimizer, acting directly upon the digital learning environment to keep the student in their Zone of Proximal Development. Its purpose is to manage the level of challenge, ensuring the student remains productive and engaged without becoming overly frustrated or disengaged.

The agent’s decision-making process is governed by a policy, $\pi$, learned through deep reinforcement learning (DRL). This policy takes the standardized student state vector, $S_t$, provided by the ontology as input and outputs an abstract action, $a_t$, designed to optimize a long-term learning objective. With an optimal policy $\pi^*$, the agent's optimal behavior is defined as $a_t^* = \pi^*(S_t)$. This optimal policy is learned by maximizing a numerical reward function. Given our application in student tutoring, this reward function can be numerically determined based on measurements of student performance, with positive rewards given for improvements in performance, engagement, or any other desired metrics.

This action, $a_t$, is a domain-agnostic vector that the client-side learning environment translates into a concrete intervention. For example, it could trigger a change in the difficulty of the next problem, provide a subtle visual hint, or unlock a helpful tool within the interface. The standardized state representation $S_t$ is crucial to decouple the agent logic from domain content and add generalizability. This allows either using a single RL agent on any topic with shared logic, or applying a method such as experience sharing from our prior work \cite{ref12, ref16} to improve the performance of multiple RL agents. The shared state vector allows for both efficient training and generalizable agent behavior depending on the needs of a specific application.

\subsection{Peer Agent}
The Peer Agent is our framework's direct answer to both the social learning gap and the educational effectiveness of LLMs. Standard LLM-powered chatbots in education are typically reactive and ungrounded; they wait for a student's query and generate a response from their general-purpose trained knowledge, making them prone to hallucinating facts. Further, they require students to have an understanding of what they do not know in order to ask appropriate questions. To overcome these issues, we designed the Peer Agent as a proactive, ontology-grounded social partner. Its intelligence is derived from two core mechanisms: proactive triggering and ontology-constrained generation.

Proactivity stems from rules embedded within the ontology that actively monitor the student's state, $S_t$. For example, a rule can be defined as "IF ($frustration > 0.8$ AND $errors > 3$) THEN $trigger(encourage\_and\_reframe)$". When this rule fires, it dispatches a trigger to the Peer Agent, prompting it to initiate a supportive, context-aware conversation without any direct student input. This solves the problem of students who are too frustrated or lost to ask for help.

Ontology-constrained generation addresses the effectiveness issue by altering the LLM's reasoning process, a method inspired by recent work in graph-constrained reasoning and retrieval-augmented generation (RAG). A typical chatbot models the probability of a response $R$ given only a query $Q$:

$$P(R|Q) = \prod_{i=1}^{|R|}P(r_i|Q, r_{<i})$$

This process is where hallucination occurs. Our approach, in contrast, tightly couples the LLM's generation to our symbolic ontology. The response is conditioned not just on the query, but on the full student state $S_t$ and, most critically, on a set of grounding knowledge and pedagogical strategies, $K_O$, retrieved from the ontology. This can be conceptualized as:

$$P(R|Q, S_t) = \prod_{i=1}^{|R|}P(r_i|Q, S_t, K_O, r_{<i})$$

where $K_O = retrieve(Ontology, Q, S_t)$. This process ensures the LLM is "thinking" with verified information. Our system retrieves relevant concepts, prerequisite relationships, and expert-defined conversational tactics from the ontology. This retrieved knowledge, $K_O$, is then used to construct a detailed prompt that constrains the LLM, steering its output to be both factually correct and pedagogically appropriate.

This neuro-symbolic approach transforms the LLM from an unreliable question-answering machine into a trustworthy partner. It directly addresses educational applications of LLMs by grounding every utterance in expert-defined knowledge, and it fills the social learning gap by enabling proactive and educationally-sound peer dialogue.

\section{Framework in Practice}\label{sec:implementation}
This section provides a practical methodology for implementing the conceptual framework. We detail how the ontology enables a "plug-and-play" approach to educational support and how the agents leverage it, using examples from two distinct games: \textit{Gridlock} \cite{ref17} (university-level digital logic) and \textit{SPARC} \cite{sparc1, sparc2} (middle-school biology).

\subsection{The Educational Ontology}
The ontology is key to our framework's generalizability. Here, we provide an overview of its structure and the processes for its creation and expansion.

\subsubsection{A Generic Template for Domain Knowledge}
To adapt the framework to a new topic, an educator populates a file following a generic template. This enforces a consistent structure that the AI agents can interpret, regardless of the subject matter. Fig. \ref{fig:template} shows a possible template for a single educational concept.

\begin{figure}[h]
	\begin{lstlisting}[breaklines=true]
{
  "concept_id": "<unique_id>",
  "display_name": "<Name>",
  "description": "<Concept_explanation>",
  "prerequisites": ["<concept_id_1>", "<concept_id_2>"],
  "difficulty_level": <float_from_0.0_to_1.0>,
  "associated_media": ["<path_to_video.mp4>"],
  "pedagogical_rules": {
    "<trigger_condition>": {
      "update_state": {"<metric>": "<operation_value>"},
      "trigger_peer": "<prompt_strategy_id>"
      }
    },
  "llm_grounding_info": {
    "key_facts": ["<Fact_1>", "<Fact_2>"],
    "common_misconceptions": ["<Misconception_1>"]
  }
}
	\end{lstlisting}
	\caption{A generic template for a single concept within the educational ontology. Placeholders in angle brackets are filled in by a domain expert.}\textbf{}
	\label{fig:template}
\end{figure}

\subsubsection{The Ontology Development Process}
Expanding the ontology or creating a new one is an iterative, expert-driven process:
\begin{enumerate}
	\item \textbf{Concept Identification:} A domain expert identifies the core concepts and learning objectives for the new topic.
	\item \textbf{Structural Definition:} The expert defines the relationships between concepts (e.g., prerequisites) and assigns attributes like difficulty.
	\item \textbf{Pedagogical Rule Authoring:} The expert defines simple, event-driven rules, specifying how the student's state should be updated or what social support should be triggered.
	\item \textbf{Population and Validation:} This information is populated into files following the template. The new ontology is then loaded and tested to ensure the agents behave as expected.
\end{enumerate}
This process allows for rapid adaptation to new domains without changing the core AI agent code.

\subsection{Standardizing Learner Data for Generalization}
A key function of the ontology is to transform disparate raw data from different learning environments into a single, standardized format.

\subsubsection{The Standardized State Vector}
The output of the transformation pipeline is the Standardized State Vector ($S_t$). This vector represents a snapshot of the student’s condition and serves as the common input for both AI agents. It is composed of a set of generic, pedagogically-grounded metrics that are domain-agnostic. While the specific metrics can be configured for a given application, our framework is designed to model a rich picture of the learner. Examples of metrics that can be included are:
\begin{itemize}
	\item \textbf{Proficiency:} The student's estimated mastery of a specific concept.
	\item \textbf{Learning Rate:} The change in proficiency over a recent period, indicating the trajectory of learning.
	\item \textbf{Frustration:} An inferred metric indicating negative affect, often triggered by repeated errors or long periods of inactivity.
	\item \textbf{Engagement:} A measure of the student's level of interaction with the material, such as time on task or use of optional resources.
	\item \textbf{Effort:} A metric reflecting the amount of work the student is putting in, such as the number of attempts on a problem.
	\item \textbf{Confidence:} The student's belief in their ability to succeed, which can be self-reported or inferred.
	\item \textbf{Exploration:} A measure of how much the student interacts with non-essential but related content or game areas.
	\item \textbf{Metacognition:} An estimate of the student's self-awareness, often inferred from help-seeking behavior or self-assessment.
\end{itemize}

\subsubsection{Transforming Disparate Data into a Standardized Format}
To illustrate how generalization is achieved, we will walk through a concrete example using the highly disparate raw data logs from \textit{Gridlock} (a structured CSV format with direct performance scores) and \textit{SPARC} (a less structured JSON event stream).

Table \ref{tab:raw_gridlock} shows a sample line from a \textit{Gridlock} log file. The key columns for our purposes are `Score`, `Time`, `Attempts`, and the `self-rated confidence` and `frustration score` values embedded within the vectors. Other values are present, but have been excluded for brevity.

\begin{table}[h]
	\caption{Example Raw Data from \textit{Gridlock} (CSV).}
	\centering
	\begin{tabular}{p{0.9\columnwidth}}
		\toprule
		\textbf{Log Entry Snippet} \\
		\midrule
		\texttt{...,(0.5, 10.05, 0.5, 0.5, ...),...,3,...} \\
		\bottomrule
	\end{tabular}
	\label{tab:raw_gridlock}
\end{table}

So, this particular student scored 0.5 out of a possible 1.0, took 10.05 seconds to complete the section, rated their confidence as middling (0.5 out of 1.0), and this is their 3rd attempt.

Table \ref{tab:raw_sparc} shows a sample of events from a \textit{SPARC} log. This data is a stream of actions, and pedagogical meaning must be inferred from the sequence and type of events.

\begin{table}[h]
	\caption{Example Raw Data Stream from \textit{SPARC} (JSON).}
	\centering
	\begin{tabular}{p{0.9\columnwidth}}
		\toprule
		\textbf{Incoming \textit{SPARC} Events} \\
		\midrule
		\texttt{\{"eventType":"DrivingPlayer...", ...\}} \\
		\texttt{\{"eventType":"ConversationStep", ...\}} \\
		\texttt{\{"eventType":"WordGameEnd", "eventData":"..."\}} \\
		\bottomrule
	\end{tabular}
	\label{tab:raw_sparc}
\end{table}

SPARC uses specific events, each with timestamps. Excerpted above, we see an event for a player driving a vehicle, completing a step of a conversation, and completing a word game. Each event includes some numerical results, such as the player's position in the world, the ID of the character they are talking with, or their submission to the word game, but these specifics are omitted here.

The system uses a set of mapping rules, guided by the ontology, to interpret this raw data and populate the Standardized State Vector. The key is that the rules are different for each game, but the output structure is identical. We illustrated several mapping processes for both \textit{Gridlock} and \textit{SPARC} below.

\begin{itemize}
	\item \textbf{To calculate Proficiency:} For \textit{Gridlock}, proficiency is mapped directly from the `Score` column (e.g., 0.5). For \textit{SPARC}, it is inferred by parsing the `eventData` of a `WordGameEnd` event to count correct vs. incorrect answers and calculating a percentage.
	\item \textbf{To calculate Effort:} For \textit{Gridlock}, effort is mapped from the `Attempts` column (e.g., 3). For \textit{SPARC}, it is inferred by counting the number of `ConversationStep` events within a single dialogue.
	\item \textbf{To calculate Frustration:} For \textit{Gridlock}, this is a direct mapping from the `frustration score` in the log's vector (e.g., 0.5). For \textit{SPARC}, it is inferred when multiple incorrect answers are logged in a `WordGameEnd` event in a short period.
	\item \textbf{To calculate Exploration:} This metric is not directly available in the \textit{Gridlock} log line, but could be inferred from the number of evaluations a student has undergone, as some evaluations in \textit{Gridlock} are optional. For \textit{SPARC}, it is calculated by counting the number of unique `DrivingPlayer...` events, which indicates the student is moving around the environment to discover new things.
\end{itemize}

This standardized vector serves as a bridge between varied learning environments and the domain-agnostic AI agents. It effectively decouples the agents' intelligent control logic from the specific data sources of any single game. By creating a consistent, structured input space, this vector is the critical enabling mechanism that allows the same machine learning models, particularly the reinforcement learning policy of the Tutor Agent, to operate, learn, and generalize their strategies across entirely different educational contexts.

\begin{lstlisting}[breaklines=true]
# Outgoing Standardized Feature Vector
student_id: "s101"
features: {
  "proficiency": 0.5,
  "learning_rate": 0.1,   # Change since last vector
  "frustration": 0.5,     # Inferred or direct
  "engagement": 0.7,      # Inferred from time/actions
  "effort": 0.6,          # Inferred from attempts
  "confidence": 0.5,      # Self-reported or inferred
  "exploration": 0.3,     # Inferred from movement
  "metacognition": 0.4    # Inferred from help-seeking
}
\end{lstlisting}

\subsection{Peer Agent}
The Peer Agent uses the ontology to generate proactive, educationally-aligned dialogue. The `trigger\_peer` value in the ontology's rules corresponds to a specific conversational strategy. Table \ref{tab:llm_examples} shows two scenarios from the \textit{SPARC} game and the resulting LLM output.

\begin{table*}[t]
	\caption{Example prompts and outputs for the LLM-powered Peer Agent in different scenarios.}
	\label{tab:llm_examples}
	\centering
	\begin{tabular}{p{0.2\textwidth} p{0.35\textwidth} p{0.35\textwidth}}
		\toprule
		\textbf{Scenario \& Trigger} & \textbf{Internal Prompt to LLM (Simplified)} & \textbf{Example Peer Agent Output} \\
		\midrule
		Student repeatedly fails to identify the Pulmonary Artery. \newline
		\textbf{Trigger:} 'encourage\_and\_reframe' & You are a friendly peer. The student is frustrated with the 'Pulmonary Artery'. \newline KEY FACT: The Pulmonary Artery is unique because it carries deoxygenated blood away from the heart. \newline MISCONCEPTION: Students often think all arteries carry oxygenated blood. \newline GOAL: Encourage them and hint at the key fact without giving the answer. & "Hey, no worries! This part is tricky for a lot of people. I remember getting stuck here too. Maybe think about what makes this artery special compared to the others. Most arteries carry blood *away* from the heart, right? What kind of blood is this one carrying?" \\
		\midrule
		Student correctly traces the entire circulatory system on the first try. \newline
		\textbf{Trigger:} 'reinforce\_and\_extend' & You are an encouraging peer. The student just mastered the circulatory system. \newline KEY FACT: The circulatory system works with the respiratory system to deliver oxygen. \newline GOAL: Congratulate them and ask a question to extend their thinking to a related topic (the lungs). & "Whoa, you got that on the first try! That's awesome! It's cool how everything is connected. It makes me wonder, where does the blood go to pick up all the oxygen in the first place?" \\
		\bottomrule
	\end{tabular}
\end{table*}

\section{Ongoing Research}
While our framework provides a path toward generalizable and effective pedagogical agents, its implementation also highlights a few challenges that can guide future research. Building effective pedagogical agents is an ongoing, iterative process to improve the synergy between humans and machines.

\subsection{Main Challenges}
Given our system's dependence on the educational ontology, the creation of the ontology requires significant resources and expert input. While it only needs to be created once, the effort required to create the ontology highlights a few challenges.
\begin{itemize}
	\item \textbf{Ontology Creation:} While the ontology enables rapid cross-domain generalization, the initial creation of a high-quality ontology remains a manual, time-consuming process that requires significant domain and pedagogical expertise \cite{ji2021survey}. This limits the speed at which the framework can be deployed to entirely new fields.
	\item \textbf{The "Cold Start" Problem for the Tutor Agent:} The RL-based Tutor Agent learns an optimal policy through interaction data. When deployed in a new domain, it begins with little to no data, forcing it to rely on a generic or random policy until it has gathered enough experience to learn effectively \cite{taylor2009transfer}. Our prior work addressed this issue through experience sharing \cite{ref12}, which led to a moderate improvement in agent performance. But appropriate behavior between different educational domains cannot be guaranteed to be identical, so the "cold start" problem remains a potential source of reduced system performance.
	\item \textbf{Nuance in State Abstraction:} The process of mapping diverse, raw logs into a single standardized state vector is a powerful abstraction, but it risks losing important, context-specific information. An action that signals frustration in one game might mean something different in another, and our current mapping rules may not capture this full nuance.
\end{itemize}

\subsection{Future Directions}
The aforementioned challenges of our current framework directly inform our agenda for future research. We are actively exploring solutions to these challenges, with a focus on creating more autonomous, intelligent, and symbiotic systems.
\begin{itemize}
	\item \textbf{Semi-Automated Ontology Generation:} To address the ontology creation bottleneck, we are investigating methods for using LLMs to assist experts in the ontology creation process. This involves developing techniques to have LLMs parse textbooks, curricula, and academic papers to generate a high-quality draft ontology, which an expert can then refine and validate \cite{pan2024unifying}.
	\item \textbf{Multi-Modal State Representations:} To create a more detailed understanding of the learner, we plan to move beyond interaction logs. Future work will focus on integrating multi-modal data streams,such as computer vision, to infer confusion from facial expressions or sentiment analysis of a student's dialogue. Ultimately, additional data modalities can still be transformed into our proposed generic data format, adding further information to improve the accuracy of our student data \cite{giannakos2023role}.
\end{itemize}

\section{Conclusion}
This article addressed a central challenge in educational AI: the trade-off between the pedagogical effectiveness of traditional, rule-based Intelligent Tutoring Systems and the conversational flexibility of modern Large Language Models. We propose a neuro-symbolic, multi-agent framework designed to provide generalizable and autonomous educational support, using multiple agents to create a synergistic system. By grounding our system in an Educational Ontology, we show the successful combination of structured, adaptive scaffolding from a reinforcement learning-optimized Tutor Agent with the safe, context-aware dialogue of an LLM-powered Peer Agent. Our framework is exemplified through case studies on two educational games to show data transformation and example responses.

Methodologically, this paper contributes a practical architecture for the design of more robust and effective human-machine systems. The proposed framework, which leverages a symbolic ontology to guide and constrain neural agents, serves as a blueprint for researchers and practitioners in education and human-machine systems. It provides a concrete example of intelligent systems that are not only effective but also adaptable and safe for real-world application.

Our future work will continue to build upon this foundation. We intend to focus on developing a more advanced, specialized version of this framework that reduces the current limitations, primarily by exploring semi-automated methods for ontology construction and enhancing the data abstraction process. Our ultimate goal is to refine this architecture, paving the way for the next generation of truly personalized, reliable, and scalable pedagogical agents.

\bibliographystyle{ieeetr}
\bibliography{refs}

\vfill
\end{document}